\documentclass{llncs}
\usepackage{algorithm}
\usepackage{algpseudocode}
\usepackage{amsmath}

\begin{document}
\title{Complexity Analysis of the Parallel Guided Ejection Search for the Pickup and Delivery Problem with Time Windows}

\author{Miroslaw Blocho \and Jakub Nalepa}

\institute{Silesian University of Technology, Gliwice, Poland\\
\email{jakub.nalepa@polsl.pl}}
\maketitle
\begin{abstract}

This paper presents the pessimistic time complexity analysis of the parallel algorithm for minimizing the fleet size in the pickup and delivery problem with time windows. We show how to estimate the pessimistic complexity step by step---this approach can be easily adopted to other parallel algorithms for solving complex transportation problems.

\end{abstract}



\section{Introduction}
The pickup and delivery problem with time windows (PDPTW) is the NP-hard problem of serving a number of transportation requests using a fleet of vehicles.
Each request is a pair of the pickup and delivery operations which must be performed in the appropriate order.
Moreover, each travel point should be visited within its time window, the size of vehicles cannot be exceeded, and all trucks should start and finish their service in the depot. The PDPTW is a hierarchical-objective discrete optimization problem---the main objective is to minimize the number of vehicles (fleet size), whereas the secondary objective is to optimize the traveled distance.

\section{Complexity Analysis}

\algblock{ParFor}{EndParFor}
\algnewcommand\algorithmicparfor{\textbf{for}}
\algnewcommand\algorithmicparto{\textbf{to}}
\algnewcommand\algorithmicpardoinparallel{\textbf{do in parallel}}
\algnewcommand\algorithmicendparfor{\textbf{end\ for}}
\algrenewtext{ParFor}[2]{\algorithmicparfor\ #1\ \algorithmicparto #2 \algorithmicpardoinparallel}
\algrenewtext{EndParFor}{\algorithmicendparfor}

\algblock{For}{EndFor}
\algrenewtext{For}[2]{\algorithmicparfor\ #1\ \algorithmicparto #2 \textbf{do}}
\algrenewtext{EndFor}{\algorithmicendparfor}

\begin{algorithm}
\begin{center}
\begin{algorithmic}[1]
\Function{ParallelRouteMinimization}{}
	\ParFor{$P_i \gets P_1$}{~$P_N$}
		\State Create an~initial solution $\sigma_{init}$; \Comment{$\theta(n)$}
		\State  $\sigma_{best}$ $\leftarrow$ $\sigma_{init}$; \Comment{$\theta(n)$}
		\State {\it finished\/} $\leftarrow$ \textbf{false}; \Comment{$O(1)$}
		\While{\textbf{not} {\it finished\/}} \Comment{max. $z_1$ iterations}
		\State Initialize ejection pool \textit{EP} with requests from a route $r$; \Comment{$O(n)$}
		\State Initialize penalty counters for each $p[h_j] := 1, j = 1, 2, ..., n$; \Comment{$\theta(n)$}
		\While{$EP$ $\neq$ $\emptyset$ \textbf{and} \textbf{not} $finished$} \Comment{max. $z_2$ iterations}
			\State Select and remove $h_{in}$ from \textit{EP} \Comment{$O(1)$}
			\If{$S_{in}^{fe}(h_{in}, \sigma) \neq \emptyset$} \Comment{$O(n^{2})$}
				\State $\sigma$ $\leftarrow$ random solution $\sigma^{\prime}\in S_{in}^{fe}(h_{in}, \sigma)$; \Comment{$O(1)$}
			\Else
				\State $\sigma$ $\leftarrow$ Squeeze($h_{in}$, $\sigma$); \Comment{$O(n^{4})$}
			\EndIf
			\If{$h_{in}$ $\notin$ $\sigma$} \Comment{$O(1)$}
			\State $p[h_{in}] := p[h_{in}] +1$; \Comment{$O(1)$}
			\For{$k \gets 1$}{~$k_{max}$}
				\State Get $S_{ej}^{fe}(h_{in}, \sigma)$ with min. $P_{sum}$ = $p[h^{(1)}_{out}]$ + $\ldots$ + $p[h^{(k)}_{out}]$; \Comment{$O(n^{k+2})$}
				\If{$S_{ej}^{fe}(h_{in}, \sigma) \neq \emptyset$} \Comment{$O(1)$}
					\State $\sigma$ $\leftarrow$ random solution $\sigma^{\prime} \in S_{ej}^{fe}(h_{in}, \sigma) \neq \emptyset$; \Comment{$O(1)$}
					\State Add the ejected cust.: $h^{(1)}_{out}, \ldots, h^{(k)}_{out}$ to the \textit{EP}; \Comment{$O(1)$}
					\State \textbf{break};
				\EndIf
			\EndFor
			\State $\sigma$ $\leftarrow$ Perturb($\sigma$); \Comment{$O(In^4)$}
			\EndIf
\State $finished \leftarrow$ Cooperate(); \Comment{$O(pn)$}
\EndWhile

\If{$EP \neq \emptyset$} \Comment($O(1)$)
	\State $\sigma_{best} \leftarrow \sigma_{init}$ ; \Comment{$\theta(n)$};
\Else
	\State $\sigma_{best} \leftarrow \sigma$ ; \Comment{$\theta(n)$};
\EndIf

\EndWhile

\While{\textbf{not} $finished$} \Comment($O(1)$)
\State $finished \leftarrow$ Cooperate(); \Comment{$O(pn)$}
\EndWhile

	\EndParFor
\State \Return best solution from all processes; \Comment{$O(1)$}
\EndFunction
\caption{The parallel algorithm for minimizing the fleet size in the PDPTW.}
\label{ParallelRMAlgorithm}
\end{algorithmic}
\end{center}
\end{algorithm}

The theoretical algorithm analysis is one of the most important issues in the computational complexity theory---it provides theoretical estimates for the time and resources needed by the algorithm solving a given computational problem. In this short paper, we estimate the time complexity of our parallel algorithm for minimizing the fleet size in the PDPTW~\cite{Nalepa20153PGCIC} in the asymptotic sense using the \textit{Big O} and \textit{Big-theta} notations. Let us assume that the parallel algorithm \textit{R} solves the problem \textit{X} of the \textit{n} data input size.
The pessimistic time complexity of \textit{R} is defined as: $T(p,n)=sup_{d \in D_n}\{t(p,d)\}$, where \textit{p} is the number of processors; $D_n$ indicates all sets of data inputs \textit{d} of \textit{n} size, $t(p,d)$ is the number of computation steps carried out for \textit{d} by all the processors.
The pessimistic time complexity $T(p,n)$ is important since it is used to estimate the speedup of the parallel algorithms $S(p,n)$:
$S(p,n)=T^{*}(1,n)/T(p,n)$, where $T^{*}(1,n)$ is the pessimistic time complexity of the fastest known sequential algorithm solving a given computational problem.
The maximum speedup is \textit{p}, as using \textit{p} processors, the total computation time can be decreased \textit{p} times (unless we face the superlinear speedup).
The cost of the parallel algorithm is finally defined as the sum of all operations carried out by all processes, and is defined as $C(p,n)=pT(p,n)$. In the complexity analysis of our parallel algorithm we assumed that \textit{n} indicates the number of requests and \textit{p} is the number of parallel processes.

The pessimistic time complexity analysis of the parallel algorithm is presented in Algorithm~\ref{ParallelRMAlgorithm}. The time complexities of simple operations such as: atomic operations (lines: 9, 10, 22, 30), setting and checking the termination condition (lines: 5, 9, 36), selecting a random solution from already created sets (lines: 12, 21), increasing penalty counters (line~17) are $O(1)$. The following operations: creating initial solution (line~3), copying solutions (lines: 4, 31, 33), ejection pool initialization (line~7), initialization of penalty counters (line~8) require $O(n)$ time. Generating the set of feasible solutions after inserting $h_{in}$ has the $O(n^{2})$ complexity (line~11). It is the result of testing the insertions of both pickup and delivery customer of $h_{in}$ at all possible positions in $\sigma$ (a single test of the feasible insertion takes only $O(1)$ time due to utilizing the forward/backward time window penalty slacks).

The pessimistic time complexity of \textit{Squeeze} (line~14) and \textit{Perturb} (line~26; \textit{I} is the number of perturbing steps) is $O(n^4)$.
It is important to note here, that the high complexity of $O(n^4)$ comes from the out-exchange moves. Out-exchange moves are executed conditionally only when out-relocate moves (of $O(n^2))$ fail. Here, we estimate the pessimistic complexity, so out-exchange moves cannot be omitted. The most exhaustive computation step is finding the best combination of ejected $k_{max}$ requests and inserting $h_{in}$ request into $\sigma$ (set $S_{ej}^{fe}(h_{in}, \sigma)$,  line~19).
The pessimistic time complexity of this step is $O^{k_{max}+2}$, but the lexicographic search applied to this step notably decreases the average time complexity.
Subsequent attempts to find the best combination using \textit{k} from 1 to $k_{max}$ additionally helps decrease the average time (lines 18-25).
Estimating average time complexities is our ongoing research and the initial experimental results showed that average complexity for $k_{max}=3$ ranges between $O(n^{3.1}$ and $O(n^{3.6})$ comparing to pessimistic $O(n^{5})$. The pessimistic time complexity of the ring co-operation step between processors is of $O(pn)$, as it is required to send, receive and replace (if needed) the current solution (lines 28, 37).  It is important to note that full solutions are being sent to the neighboring processors in the ring only if the number of routes of the current solution decreased from previous co-operation. The complexity of this step grows linearly with number of processors $p$. However, it was shown that this negative impact is compensated by the quality of the retrieved results (the larger number of parallel processes, the better solutions).

Let $z_1$ and $z_2$ be the number of executions of the \textit{while} instructions in lines 6 and 9, respectively.
Then, the total pessimistic time complexity of the parallel algorithm is:
\begin{equation}
  T_{pes}(p,n)=s_1n+z_1(s_2n+z_2T^{in}_{pes}(p,n)+s_3n)+s_4pn,
\end{equation}
where $s_1n$, $s_2n$, $s_3n$, $s_4pn$ are costs of operations in lines 3-5, 7-8, 30-34, 36-38 for certain constants $s_1$, $s_2$, $s_3$, $s_4$, and $T^{in}_{pes}(p,n)$ is the time complexity of a single run of the inner \textit{while} in lines 9-29.
$T^{in}_{pes}(p,n)$ is calculated as:
\begin{equation}
  T^{in}_{pes}(p,n)=s_5n^4+s_6pn^{k_{max}+2} + s_7pn,
\end{equation}
where $s_5n^4$, $s_6pn^{k_{max}+2}$, $s_7pn$ are costs of operations included in lines 10-15, 16-27, 28, for constants $s_5$, $s_6$, $s_7$. Hence, the full pessimistic time complexity of the parallel algorithm becomes:
\begin{equation}
\begin{aligned}
  T_{pes}(p,n)=&s_1n+z_1(s_2n+z_2(s_5n^4+s_6pn^{k_{max}+2} + \\&s_7pn)+s_3n)+s_4pn = O(n^{k_{max}+2} + pn).
  \end{aligned}
\end{equation}

Although this pessimistic time complexity appears significant, the parallel algorithm was shown to execute (and converge) very fast in practice~\cite{Nalepa20153PGCIC}.

\section{Conclusion and Outlook}

In this short paper, we analyzed the theoretical pessimistic time complexity of our parallel algorithm for minimizing the number of routes in the PDPTW. We showed how to estimate the pessimistic complexity step by step---this approach can be easily adopted to other parallel algorithms for solving complex transportation problems. Our ongoing research focuses on estimating (theoretically) the time complexity of our parallel algorithm for minimizing the total distance traveled in the PDPTW. Also, we plan to investigate the average time and memory complexities of both parallel algorithms.

\section*{Acknowledgment}
This research was supported by the National Science Centre under research Grant No. DEC-2013/09/N/ST6/03461.

\bibliographystyle{abbrv}
\bibliography{ref_all_jn}

\end{document}